\documentclass[showpacs,preprintnumbers,prd,nofootinbib,floats,amssymb,floatfix]{revtex4}
\usepackage{amsmath}
\usepackage{amsfonts}
\usepackage{graphicx}
\usepackage{hyperref}
\usepackage{amsmath}
\usepackage{amstext}
 
\begin{document}
\title{ Faddeev-Jackiw   quantization of an Abelian and non-Abelian exotic action for gravity in three dimensions}
 \author{ Alberto Escalante } \email{aescalan@ifuap.buap.mx}
\author{J. Manuel-Cabrera } \email{jmanuel@ifuap.buap.mx}
 \affiliation{  Instituto de F{\'i}sica, Benem\'erita Universidad Aut\'onoma de Puebla, \\
 Apartado Postal J-48 72570, Puebla Pue., M\'exico, }
\begin{abstract}
A detailed Faddeev-Jackiw quantization of an Abelian and non-Abelian exotic action for gravity in three dimensions is performed.  We obtain for the theories under study  the constraints,   the gauge transformations, the generalized Faddeev-Jackiw brackets and  we perform the counting of physical  degrees of freedom. In addition,  we  compare our results with those found in the literature where the canonical analysis  is developed, in particular,  we  show that both the generalized   Faddeev-Jackiw brackets  and  Dirac's brackets coincide to each other. Finally we discuss some remarks and prospects.
\end{abstract}
 \date{\today}
\preprint{}
\maketitle
\section{INTRODUCTION}
Nowadays, the study of singular systems becomes to be   an important research subject. In fact, the gravitational interaction described by means Palatini's  action,  the physics of the fundamental interactions based  on the standard model, Maxwell theory, Yang-Mills theories and  BF theories  \cite{1a, 2a, 3a} etc.,  are several  examples of important singular systems,  and its study  at classical and quantum level becomes to be mandatory.  In fact, the study of   those dynamical systems has been developed  through their   symmetries and these symmetries form part of a relevant  information in both  the classical and quantum context. In this respect,  it is well-known  that there is  a powerful  formalism for studying the symmetries of singular systems, in particular those singular systems having  an important symmetry called gauge symmetry,  that formalism is know as the   Dirac-Bergman  method for  constrained systems \cite{4a}. Dirac's canonical formalism is an elegant approach for obtaining relevant physical information of a  theory, namely, the identification of the  physical degrees of freedom, the gauge transformations,  the complete structure of the constraints and the obtention of the extended  action,  all this information is useful   because a strict study of the symmetries will allow us to have a   guideline to make the best progress in the quantization. However,  if  a pure Dirac's  canonical analysis is performed, in general    it is complicated to develop  the classification of  the constraints in  first  and second class \cite{5a, 6a, 7a};  the classification of the constraints is an important step to perform because first class constraints are generators of gauge transformations and allow us identify observables. Moreover,  second class constraints allow  us to construct the Dirac brackets and also they are useful for identifying  the  Lagrange multipliers. Therefore,  to develop    all the  steps of a pure canonical analysis becomes to be   mandatory but it  is not easy to perform \cite{8aa}.\\
On the other hand, there is an alternative method   for studying  singular systems  in a different way,    that  is  the  well-known  Faddeev-Jackiw  [FJ]  formalism \cite{1}. The [FJ]  framework is a symplectic description of constrained quantization,  where the degrees of freedom are  identified by means  the so-called  symplectic variables. In fact, for studying any theory  we can choose as  symplectic variables either  configuration space or  the phase space;  in [FJ] framework there is a freedom for choosing the symplectic variables. Furthermore,  as the system under study is singular  there will  be constraints and the   [FJ] approach has the  advantage that  all the  constraints of the theory are at the same footing, namely,   it is not necessary  perform  the  classification of the constraints in primary, secondary, first class or second class  such as in Dirac's method is done. Moreover,   in [FJ] approach   also it is possible to obtain the gauge transformations of the theory and the generalized [FJ]  brackets  coincide  with the Dirac's  ones. The [FJ] framework has been applied to systems such as [YM] theory \cite{13}, QCD theory \cite{14b}, first order Wess-Zumino terms \cite{15b},   theories with extra dimensions \cite{16b} and several others interesting  systems \cite{22a}. We can  observe in all  those works that the [FJ] approach is an  elegant  alternative   for analysing gauge systems with certain  advantages lacking    the Dirac approach.  \\
Because of  the explained above, in this paper we perform both the Dirac  and [FJ] analysis  to an exotic Abelian theory  and non-Abelian exotic action describing gravity in three dimensions. In fact, for the former   we will show that if in Dirac's formalism we perform the analysis without fixing the gauge and we eliminate  only the second class constraints through Dirac's brackets and  remaining the first class ones, then  it is possible  reproduce those  Dirac's results by working in [FJ] with the configuration space as symplectic variables,  and using the temporal gauge  in order to invert the symplectic matrix. Moreover, if in  Dirac's method we  perform the analysis and we  fixing the gauge converting  the first class constraints in second class and   we construct the corresponding Dirac's brackets, then  in [FJ] framework it is possible to reproduce these  Dirac's results,  but now we will  work   by using the phase space as symplectic variables;  in order to invert the symplectic matrix  we will  fix   the gauge using the   Coulomb gauge. In addition, we will show that if  in Dirac's method  fixing or not the gauge,  the  constructed  Dirac's  brackets and the generalized [FJ] brackets coincide to each other. Furthermore, we will extend our analysis for  a non-Abelian exotic theory describing gravity   and   we will reproduce in an  elegant way by means [FJ]  the results reported in \cite{9} where the Dirac approach was  performed, all these important results will be clarified along the paper. \\
The paper is organized as follows, in Sect. II  we will  perform the [FJ] analysis for  an Abelian exotic action  in three dimensions,   we will  obtain the constraints of the theory, the gauge transformations and  we will carry out the counting of physical degrees of freedom concluding that the theory under study is a  topological one. In addition, we will reproduce the results obtained from the canonical analysis  where the second class constraints are eliminated by introducing the Dirac brackets and we show that the generalized [FJ] brackets coincide with those  Dirac's  brackets. Then,  in Dirac's method we will fix the gauge  converting  the first class constraints in second class and again,   the Dirac  brackets will be constructed, thus, in order to reproduce these  results  by using  [FJ],   we will perform our  analysis by working now  with  the phase space as symplectic variables and we will show the equivalence between  the generalized [FJ]  and Dirac's brackets.  In Sec. III we will extend our analysis developed  in previous  sections by performing  the [FJ] analysis for the  exotic action describing gravity, in particular, we will reproduce all  the Dirac results  reported in \cite{9} where a pure canonical analysis was reported. In Sec. IV we provide a summary and prospects. \\
\\
\section{Faddeev-Jackiw quantization of an Abelian exotic action}
It is well-known  that in three dimensions there is an alternative action to Palatini's Lagrangian  reproducing  Einstein's equations  with  cosmological constant,  that action is called  exotic action for gravity. Exotic action for gravity can be seen as  a limit of   other  theories  such as topological gravity with torsion  or topologically massive gravity \cite{9a, 9b, 9, 15c, 16c}, the action is given by 
\begin{eqnarray}
S^{Exotic}[A,e]&=&\int_M A^{I} \wedge d A_{I} + \frac{1}{3}\epsilon_{IJK}A^{I}\wedge A^{J} \wedge A^{K}\nonumber\\ &\,& + \Lambda\int_M  e^{I}\wedge d_{A}e_{J}.
\label{1}
\end{eqnarray}
here,  $\Lambda$ is the cosmological constant,  $A^{I}=A^{I}_{\mu}dx^{\mu}$ is the one-form in  the adjoint representation of the Lie algebra of  $SO(2, 1) $ \cite{15c, 16c},  $e^{I}$ corresponds to the dreibein  field, $\mu,\nu=0,1,2$ are spacetime indices, $x^{\mu}$ are the coordinates that label the points for the 3-dimensional spacetime manifold M and $I,J=0,1,2$ are internal indices that can be raised and lowered by the  internal metric $\eta_{IJ}=diag(-1,1,1)$.  From (\ref{1}) we can identify the Lagrangian density of an abelian exotic theory, this is
\begin{equation}
\mathcal{L}=\frac{1}{4}\epsilon^{\mu\nu\lambda}\left (A^{I}_{\mu}F_{I \nu \lambda}+\Lambda e^{I}_{\mu}( \partial_{\nu}e_{I \lambda }- \partial_{\lambda}e_{I \nu })\right),
\label{eq1}
\end{equation}
where, $A^I_{\mu}$ is a set of three  $U(1)$  gauge potentials,  $e^I_{\mu}$ is the "frame" field and  $F^I_{ \nu \lambda}= \partial_\nu A^I_\lambda- \partial_\lambda A^I_\nu$ is the strength field. In this manner, we will develop  the [FJ] analysis  to the action (\ref{eq1}). For this aim, we  perform the 2+1 decomposition  and we identify  the first order  symplectic Lagrangian given by
\begin{equation}
\mathcal{L}^{(0)}=\frac{1}{2}\epsilon^{0ij}(A^{I}_{j}\dot{A}_{iI}+\Lambda e^{I}_{j}\dot{e}_{iI})-V^{(0)},
\label{eq2}
\end{equation}
where $V^{(0)}=-\frac{1}{2}\epsilon^{0ij}(A^{I}_{0}F_{ijI}+2\Lambda e^{I}_{0}\partial_{i} e_{jI})$. The corresponding symplectic equations of motion are given by \cite{1}
\begin{equation}
f^{(0)}_{ij}\dot{\xi}^{j}=\frac{\partial V^{(0)}(\xi)}{\partial\xi^{i}},
\label{eq4}
\end{equation}
where the symplectic matrix $f^{(0)}_{ij}$ takes the form
\begin{equation}
f^{(0)}_{ij}(x,y)=\frac{\delta a_{j}(y)}{\delta\xi^{i}(x)}-\frac{\delta a_{i}(x)}{\delta\xi^{j}(y)},
\label{4}
\end{equation}
with $\xi{^{(0)i}}$ and $a{^{(0)}}{_{i}}$ representing  a set of symplectic variables. It is important to comment, that in [FJ] framework we are free for choosing the symplectic variables; we can choose either the   configuration variables or the phase space variables. In fact, in order to  obtain by means a different way the  different  scenarios    applying the Dirac  method,   in this paper we will work with both. In this respect,  let us reproduce by means the  [FJ] method  the results found   in the Appendix A (see the subsection A),  where a pure  canonical analysis was developed for  the action (\ref{1}) and  we have found the Dirac brackets by eliminating only the second class constraints    and remaining the first class ones. So, for this aim  we observe   from the symplectic Lagrangian (\ref{eq2}) that  it is possible identify  the following  symplectic variables    $ \xi{^{(0)i}}(x)=\{e^{I}_{i},e^{I}_{0},A^{I}_{i},A^{I}_{0}\}$  and the components of the symplectic 1-forms are  $a{^{(0)}}{_{i}}(x)=\{\frac{1}{2}\Lambda\epsilon^{0ij}e_{Ij},0,\frac{1}{2}\epsilon^{0ij}A_{Ij},0\}$. Hence, by using our set of symplectic variables,  the symplectic matrix (\ref{4}) takes the form
\begin{eqnarray}
f^{(0)}_{ij}(x,y)=
\left(
 \begin{array}{cccc}
   -\Lambda\epsilon^{0ij}\eta_{IJ}&0&0&0\\
   0&0&0&0\\
   0&0&-\epsilon^{0ij}\eta_{IJ}&0\\
   0&0&0&0\\
  \end{array}
\right)\delta^2(x-y),
\label{eq7}
\end{eqnarray}
we observe that this matrix is  singular. In fact, in [FJ] method  this means that there are  present constraints  \cite{8, 11}. In order to obtain these constraints, we calculate  the zero modes of the symplectic matrix, the modes are given by  $(v_{i}^{(0)})_{1}^{T}=(0,v^{e^{I}_{0}},0,0)$ and $(v_{i}^{(0)})_{2}^{T}=(0,0,0,v^{A^{I}_{0}})$, where $v^{e^{I}_{0}}$ and $v^{A^{I}_{0}}$ are  arbitrary functions. In this manner, by using the zero-modes and the symplectic potential $V^{(0)}$ we can get the following constraints 
\begin{eqnarray}
\Omega^{(0)}_{I}&=&\int d^{2}x(v^{(0)})^{T}_{i}(x)\frac{\delta}{\delta\xi^{(0)i}(x)}\int d^{2}y V^{(0)}(\xi) \nonumber \\
            &=& \int d^{2}x v^{e^{I}_{0}}(x)[-\Lambda\epsilon^{0ij}\eta_{IJ}\partial_{i}e{_{j}}^{J}] \rightarrow [-\Lambda\epsilon^{0ij}\eta_{IJ}\partial_{i}e{_{j}}^{J}]=0,
    \label{eq8}
\end{eqnarray}
\begin{eqnarray}
\beta^{(0)}_{I}&=&\int d^{2}x(v^{(0)})^{T}_{i}(x)\frac{\delta}{\delta\xi^{(0)i}(x)}\int d^{2}y V^{(0)}(\xi) \nonumber \\
            &=& \int d^{2}x v^{A_{0}}(x)[-\frac{1}{2}\epsilon^{0ij}\eta_{IJ}\left[F^{J}{_{ij}}\right] \rightarrow[-\frac{1}{2}\epsilon^{0ij}\eta_{IJ}\left[F^{J}{_{ij}}\right]=0.
          \label{eq9}
\end{eqnarray}
Now,  we will  observe  if there are present more constraints in the context of [FJ]. For this aim, we write in matrix form the following system \cite{8}
\begin{equation}
   f^{(1)}_{kj}\dot{\xi}^{j}=Z_{k}(\xi),
\label{eq12}
\end{equation}
where
\begin{eqnarray}
Z_{k}(\xi)=
\left(
 \begin{array}{cccc}
   \frac{\partial V^{(0)}(\xi)}{\partial \xi^{i}}\\
   0\\
   0\\
  \end{array}
\right),
\label{eq14}
\end{eqnarray}
and
\begin{eqnarray}
f^{(1)}_{kj}=
\left(
 \begin{array}{cccc}
   f^{(0)}_{ij}\\
   \frac{\partial\Omega^{(0)}}{\partial\xi^{i}}\\
   \frac{\partial\beta^{(0)}}{\partial\xi^{i}}\\
  \end{array}
\right)=\left(
 \begin{array}{cccc}
   -\Lambda\epsilon^{0ij}\eta_{IJ}&0&0&0\\
   0&0&0&0\\
   0&0&\epsilon^{0ij}\eta_{IJ}&0\\
   0&0&0&0\\
   \Lambda\epsilon^{0ij}\eta_{IJ}\partial_{i}&0&0&0\\
   0&0&\epsilon^{0ij}\eta_{IJ}\partial_{i}&0\\
  \end{array}
\right)\delta^2(x-y).
\label{eq13}
\end{eqnarray}
We can observe that the  matrix (\ref{eq13})  is  not a square matrix as expected, however, it has   linearly independent modes given by $(v^{(1)})_{1}^{T}=(\partial_{i}v^{\lambda},v^{e^{I}_{0}},0,0,v^{\lambda},0)$ and $(v^{(1)})_{2}^{T}=(0,0,\partial_{i}v^{\alpha},v^{A^{I}_{0}},0,v^{\alpha})$. These modes are used in order to obtain more constraints. In fact,  by calculating the following contraction \cite{8}
\begin{equation}
(v^{(1)})_{k}^{T}Z_{k}=0,
\label{eq15}
\end{equation}
where $k=1,2$, we obtain that (\ref{eq15}) is an identity; thus, in the  [FJ] context  there are not more constraints for the theory under study. \\
Now, we will construct a new symplectic Lagrangian containing the information of the constraints obtained in (\ref{eq8}) and (\ref{eq9}). In order to archive this aim, we introduce to $e^{I}_{0}=\dot{\lambda}^{I}$ and  $A^{I}_{0}=\dot{\theta}^{I}$ as  Lagrange multipliers associated to those constraints, thus,  we obtain the following symplectic Lagrangian
\begin{equation}
\mathcal{L}^{(1)}=\frac{1}{2}\epsilon^{0ij}A_{jI}\dot{A}^{I}_{i}+\frac{\Lambda}{2}\epsilon^{0ij}e_{jI}\dot{e}^{I}_{i}+(\Lambda\epsilon^{0ij}\partial_{i}e_{jI})\dot{\lambda}^{I}+(\epsilon^{0ij}\partial_{i}A_{jI})\dot{\theta}^{I}-V^{(1)},
\label{eq18}
\end{equation}
where $V^{(1)}=V^{(0)}\mid_{\Omega^{(0)}_{I}=0,\beta^{(0)}_{I}=0}=0$, the symplectic potential vanish  reflecting the general covariance of the theory just like it is present in  General Relativity. In this manner, from (\ref{eq18}) we identify the following new symplectic variables $\xi{^{(1)i}}(x)=\{e^{I}_{i},\lambda^{I},A^{I}_{i},\theta^{I}\}$ and the new symplectic 1-forms
$a{^{(1)}}{_{i}}(x)=\{\frac{1}{2}\Lambda\epsilon^{0ij}e_{Ij},\Lambda\epsilon^{0ij}\partial_{i}e_{jI},\frac{1}{2}\epsilon^{0ij}A_{Ij},\epsilon^{0ij}\partial_{i}A_{jI}\}$. Hence, by using the new symplectic variables and 1-forms, we can calculate the following symplectic matrix
\begin{eqnarray}
f^{(1)}_{ij}(x,y)=
\left(
 \begin{array}{cccc}
   -\Lambda\epsilon^{0ij}\eta_{IJ}&-\Lambda\epsilon^{0ij}\eta_{IJ}\partial_{j}&0&0\\
   \Lambda\epsilon^{0ji}\eta_{IJ}\partial_{i}&0&0&0\\
   0&0&-\epsilon^{0ij}\eta_{IJ}&-\epsilon^{0ij}\eta_{IJ}\partial_{j}\\
  0&0&\epsilon^{0ji}\eta_{IJ}\partial_{i}&0\\
  \end{array}
\right)\delta^2(x-y).
\label{eq22}
\end{eqnarray}
This matrix  is still singular. However, we have showed that there are not more constraints; the non invertibility of (\ref{eq22}) means that the theory has a gauge symmetry. In this manner, we choose  the following (gauge conditions) constraints
\begin{eqnarray}
e^{I}_{0}&=&0,\nonumber\\
A^{I}_{0}&=&0,
\label{eq23}
\end{eqnarray}
which means  that $\lambda^I$ and $\theta^I$ are constants. Hence,  we construct a new symplectic Lagrangian by adding  the constraints (\ref{eq23}) with the following  $\phi_{I}$ and $\alpha_{I}$ Lagrange multipliers,   obtaining
\begin{eqnarray}
\mathcal{L}^{(2)}=\frac{1}{2}\epsilon^{0ij}\eta_{IJ}A^{I}_{j}\partial_{0}A^{J}_{i}+\frac{1}{2}\Lambda\epsilon^{0ij}\eta_{IJ} e^{I}_{j}\partial_{0}e^{J}_{i}+(\Omega^{(0)}_{I}+\phi_{I})\dot{\lambda}^{I}+(\beta^{(0)}_{I}+\alpha_{I})\dot{\theta}^{I},
\label{eq24}
\end{eqnarray}
where we can identify the following set of  symplectic variables $\xi{^{(2)i}}(x)=\{e^{I}_{i},\lambda^{I},A^{I}_{i},\theta^{I},\phi_{I},\alpha_{I}\}$ and  the 1-forms are given by $a{^{(2)}}{_{i}}(x)=\{\frac{1}{2}\Lambda\epsilon^{0ij}e_{Ij},\Omega^{(0)}_{I}+\phi_{I},\frac{1}{2}\epsilon^{0ij}A_{Ij},\beta^{(0)}_{I}+\alpha_{I},0,0\}$. By using this new set of symplectic variables, we obtain the following 24$\times$24 symplectic matrix
\begin{eqnarray}
f^{(2)}_{ij}(x,y)=
\left(
 \begin{array}{cccccccc}
   -\Lambda\epsilon^{0ij}\eta_{IJ}&\Lambda\epsilon^{0ij}\eta_{IJ}\partial_{j}&0&0&0&0\\
  \Lambda\epsilon^{0ji}\eta_{IJ}\partial_{i}&0&0&0&-\delta{^{J}}_{I}&0\\
   0&0&-\epsilon^{0ij}\eta_{IJ}&\epsilon^{0ij}\eta_{IJ}\partial_{j}&0&0\\
   0&0&\epsilon^{0ji}\eta_{IJ}\partial_{i}&0&0&-\delta{^{J}}_{I}\\
   0&\delta{^{I}}_{j}&0&0&0&0\\
   0&0&0&\delta{^{I}}_{J}&0&0\\
  \end{array}
\right)\delta^2(x-y),
\label{eq26}
\end{eqnarray}
we observe that $f^{(2)}_{ij}$ is not  singular, hence, it is  an invertible matrix.  The inverse of the matrix (\ref{eq26}) is given by the following 24$\times$24 matrix 
\begin{eqnarray}
[f^{(2)}_{ij}(x,y)]^{-1}=
\left(
 \begin{array}{cccccc}
   \frac{1}{\Lambda}\epsilon_{0ij}\eta^{IJ}&0&0&0&\delta^{I}_{J}\partial_{i}&0\\
   0&0&0&0&\delta^{I}_{J}&0\\
0&0&\epsilon_{0ij}\eta^{IJ}&0&0&\delta^{I}_{J}\partial_{i}\\
   0&0&0&0&0&\delta^{I}_{J}\\
 \delta^{J}_{I}\partial_{j}&-\delta^{J}_{I}&0&0&0&0\\
 0&0&\delta^{J}_{I}\partial_{j}&-\delta^{J}_{I}&0&0\\
  \end{array}
\right)\delta^2(x-y).
\label{eq27}
\end{eqnarray}
Therefore, from (\ref{eq27}) it is possible to identify the following [FJ]  generalized brackets given by 
\begin{eqnarray}
\{\xi_{i}^{(2)}(x),\xi_{j}^{(2)}(y)\}_{FD}=[f^{(2)}_{ij}(x,y)]^{-1},
\label{eq28}
\end{eqnarray}
thus
\begin{eqnarray}
\{e^{I}_{i}(x),e^{J}_{j}(y)\}_{FD}=[f^{(2)}_{11}(x,y)]^{-1}=\frac{1}{\Lambda}\epsilon_{0ij}\eta^{IJ}\delta^{2}(x-y),
\label{eq30}
\end{eqnarray}

\begin{eqnarray}
\{A^{I}_{i}(x),A^{J}_{j}(y)\}_{FD}=[f^{(2)}_{33}(x,y)]^{-1}=\epsilon_{0ij}\eta^{IJ}\delta^{2}(x-y),
\label{eq31}
\end{eqnarray}

\begin{eqnarray}
\{e^{I}_{i}(x),\phi_{J}(y)\}_{FD}=[f^{(2)}_{15}(x,y)]^{-1}=\delta^{I}_{J}\partial_{i}\delta^{2}(x-y),
\label{eq32}
\end{eqnarray}

\begin{eqnarray}
\{A^{I}_{i}(x),\alpha_{J}(y)\}_{FD}=[f^{(2)}_{36}(x,y)]^{-1}=\delta^{I}_{J}\partial_{i}\delta^{2}(x-y),
\label{eq33}
\end{eqnarray}

\begin{eqnarray}
\{\lambda^{I}(x),\phi_{J}(y)\}_{FD}=[f^{(2)}_{25}(x,y)]^{-1}=\delta^{I}_{J}\delta^{2}(x-y),
\label{eq34}
\end{eqnarray}

\begin{eqnarray}
\{\theta^{I}(x),\alpha_{J}(y)\}_{FD}=[f^{(2)}_{46}(x,y)]^{-1}=\delta^{I}_{J}\delta^{2}(x-y),
\label{eq35}
\end{eqnarray}
we observe that the generalized [FJ] brackets are equivalent with those given in appendix A ( see the Eqs. from   (\ref{125}) to (\ref{132})). In fact,  if we consider  in the Dirac  brackets  (\ref{125})-(\ref{132}) that  the second class constraints  (\ref{134n}) are  strongly identities, then  the Dirac brackets will correspond     to the  generalized [FJ] brackets found above. \\
Furthermore, we will find  the gauge transformations of the theory. We have seen that   (\ref{eq15})  allowed us to know  if there are more constraints in the system \cite{8}. If no new constraints arise from  (\ref{eq15}), then the  zero modes of the matrix (\ref{eq22}) will give rise to gauge transformations. In fact,  let us obtain the gauge transformations of the theory, for this aim we rewrite the Lagrangian (\ref{eq18}) in the following form
\begin{eqnarray*}
\mathcal{L}^{(1)}=\bar{a}^{(0)}_{i}\dot{\bar{\xi}}^{(0)i}+\dot{\gamma}^{\alpha}\Phi^{(0)}_{\alpha}-V^{(1)}, 
\end{eqnarray*}
where the symplectic variables set $\bar{\xi}^{(0)i}=(e^{I}_{i},A^{I}_{i})$,  $\gamma^{\alpha}=(\gamma^{1}=\lambda^{I},\gamma^{2}=\theta^{I})$, $\Phi^{(0)}_{\alpha}=(\Phi^{(0)}_{1}=\Omega^{(0)}_{I},\Phi^{(0)}_{2}=\beta^{(0)}_{I})$.  By using the symplectic variables we can construct a nonsingular matrix, namely  $\bar{f}_{ij}=\frac{\partial \bar{a}_{j}}{\partial\xi^{i}}-\frac{\partial \bar{a}_{i}}{\partial\xi^{j}}$, constructed out  with the symplectic  1-form  $\bar{a}_{i}=(\frac{1}{2}\Lambda\epsilon^{0ij}e_{Ij},\frac{1}{2}\epsilon^{0ij}A_{Ij})$. Hence, in terms of the $\bar{\xi}'s$ and the constraints we construct the following symplectic matrix 
\begin{eqnarray*}
\mathfrak{f}^{(1)}_{ij}(x,y)=
\left(
 \begin{array}{cc}
   \bar{f}&(\frac{\partial\Phi^{(0)}}{\partial\xi})\\
   -(\frac{\partial\Phi^{(0)}}{\partial\xi})^{T}&0\\
 \end{array}
\right)\delta^2(x-y),
\end{eqnarray*}
where 
\begin{eqnarray*}
\Bigg(\frac{\partial\Phi^{(0)}}{\partial \xi}\Bigg)_{i\alpha}=
\left(
 \begin{array}{cc}
 \frac{\partial\Phi^{(0)}_{1}}{\partial\xi^{1}}&\frac{\partial\Phi^{(0)}_{2}}{\partial\xi^{1}}\\
 \frac{\partial\Phi^{(0)}_{1}}{\partial\xi^{2}}&\frac{\partial\Phi^{(0)}_{2}}{\partial\xi^{2}}\\
 \frac{\partial\Phi^{(0)}_{1}}{\partial\xi^{3}}&\frac{\partial\Phi^{(0)}_{2}}{\partial\xi^{3}}\\
 \frac{\partial\Phi^{(0)}_{1}}{\partial\xi^{4}}&\frac{\partial\Phi^{(0)}_{2}}{\partial\xi^{4}}\\
 \end{array}
\right).
\end{eqnarray*}
It is easy to observe that the symplectic matrix $\mathfrak{f}^{(1)}_{ij}$ has zero-modes with the following structure \cite{11}
\begin{eqnarray}
v_{i \alpha}=
\left(
 \begin{array}{cc}
   (\bar{f}_{ij})(\frac{\partial\Phi^{(0)}_{\alpha}}{\partial\xi^{j}})\\
   1_{(\alpha)}\\
 \end{array}
\right).
\label{vecnull}
\end{eqnarray}
In the case of  gauge theories, the symplectic matrix will be  non-invertible, however, the  null eigenvectors of that matrix are generators of the intrinsic gauge symmetry. In fact, the gauge transformation of the theory are given by  \cite{11}
\begin{eqnarray}
\delta \bar{\xi}^{i}&=&(\bar{f}_{ij})^{-1}\frac{\partial \Phi^{(0)}_{\alpha}}{\partial \xi^{j}}\epsilon^{\alpha}, \nonumber\\
\delta \gamma^{\alpha}&=&\epsilon^{\alpha}, 
\label{gaugetran}
\end{eqnarray}
thus,  by using (\ref{vecnull}) we can calculate the zero-modes of the matrix (\ref{eq22}) and they are  given by $(\mathrm{w}^{(1)})_{1}^{T}=(\partial_{i}\varepsilon ^{I},\varepsilon^{I},0,0)$ and $(\mathrm{w}^{(1)})_{2}^{T}=(0,0,\partial_{i}\zeta ^{I},\zeta^{I})$. In this manner,  from (\ref{gaugetran}) the  gauge transformations are 
\begin{eqnarray*}
\delta e^{I}_{i}=\partial_{i}\epsilon^{I}, \nonumber\\
\delta e^{I}_{0}=\dot{\epsilon}^{I}, \nonumber\\
\delta A^{I}_{i}=\partial_{i}\zeta^{I}, \nonumber\\
\delta A^{I}_{0}=\dot{\zeta}^{I}, \nonumber
\end{eqnarray*}
here $\epsilon$ and $\zeta$  form a set of infinitesimal parameters characterising  the transformations.  Therefore, we can observe that the  zero-modes display the well known Abelian gauge symmetry of the model. \\
On the other hand, in the Appendix A (see the part B)  by using the Dirac method,  we have fixed the gauge in oder to convert the first class constraints in second class constraints  and we have calculated  the corresponding Dirac's brackets among physical fields and  they are given from  (\ref{133}) to (\ref{138}), thus,  in the follow section  we will  obtain by  a  different way   those results by using the  [FJ] framework. However,   let us  continue by working   with the configuration space and  now  we will fix the  following gauge 
\begin{eqnarray}
\partial^{i}e^{I}_{i}=0,\nonumber\\
\partial^{i}A^{I}_{i}=0,
\label{25v}
\end{eqnarray}
 by using   this  gauge   with its corresponding  Lagrange multipliers, namely,  ${\rho}_{I}$ and ${\gamma}_{I}$,   the symplectic  Lagrangian (\ref{eq18})  is given by
\begin{eqnarray*}
\mathcal{L}^{(2)}=\frac{\epsilon^{0ij}}{2}\eta_{IJ}A^{I}_{j}\partial_{0}A^{J}_{i}+\frac{\Lambda}{2}\epsilon^{0ij}\eta_{IJ} e^{I}_{j}\partial_{0}e^{J}_{i}+\Omega^{(0)}_{I}\dot{\lambda}^{I}+\partial^{i}e^{I}_{i}\dot{\rho}_{I}+\beta^{(0)}_{I}\dot{\theta}^{I}+\partial^{i}A^{I}_{i}\dot{\gamma}_{I},
\end{eqnarray*}
where we can choose the symplectic variables as  $\xi{^{(2)i}}(x)=\{e^{I}_{i},\lambda^{I},A^{I}_{i},\theta^{I},\rho_{I},\gamma_{I}\}$ and the 1- form $a{^{(2)}}{_{i}}(x)=\{\frac{1}{2}\Lambda\epsilon^{0ij}e_{Ij},\Omega^{(0)}_{I},\frac{1}{2}\epsilon^{0ij}A_{Ij},\beta^{(0)}_{I},\partial^{i}e^{I}_{i},\partial^{i}A^{I}_{i}\}$. Thus,  by using these symplectic variables we obtain the symplectic matrix
\begin{eqnarray*}
f^{(2)}_{ij}(x,y)=
\left(
 \begin{array}{cccccccc}
   -\Lambda\epsilon^{0ij}\eta_{IJ}&\Lambda\epsilon^{0ij}\eta_{IJ}\partial_{j}&0&0&-\delta^{J}_{I}\partial^{i}&0\\
  \Lambda\epsilon^{0ji}\eta_{IJ}\partial_{i}&0&0&0&0&0\\
   0&0&-\epsilon^{0ij}\eta_{IJ}&\epsilon^{0ij}\eta_{IJ}\partial_{j}&0&-\delta^{J}_{I}\partial^{i}\\
   0&0&\epsilon^{0ji}\eta_{IJ}\partial_{i}&0&0&0\\
   -\delta^{I}_{J}\partial^{j}&0&0&0&0&0\\
   0&0&-\delta^{I}_{J}\partial^{j}&0&0&0\\
  \end{array}
\right)\delta^2(x-y),
\end{eqnarray*}
we observe that this matrix is not singular, therefore, it is a invertible matrix. The inverse matrix is given by
\begin{eqnarray}
[f^{(2)}_{ij}(x,y)]^{-1}=
\left(
 \begin{array}{cccccc}
   0&\epsilon_{ij}\frac{\eta^{IJ}}{\Lambda}\frac{\partial^{j}}{\nabla^{2}}&0&0&-\delta^{J}_{I}\frac{\partial_{i}}{\nabla^{2}}&0\\
   \epsilon_{ji}\frac{\eta^{IJ}}{\Lambda}\frac{\partial^{i}}{\nabla^{2}}&0&0&0&\delta^{J}_{I}\frac{1}{\nabla^{2}}&0\\
0&0&0&\epsilon_{ij}\eta^{IJ}\frac{\partial^{j}}{\nabla^{2}}&0&-\delta^{J}_{I}\frac{\partial_{i}}{\nabla^{2}}\\
   0&0&\epsilon_{ji}\eta^{IJ}\frac{\partial^{i}}{\nabla^{2}}&0&0&\delta^{J}_{I}\frac{1}{\nabla^{2}}\\
   -\delta^{I}_{J}\frac{\partial_{j}}{\nabla^{2}}&-\delta^{I}_{J}\frac{1}{\nabla^{2}}&0&0&0&0\\
 0&0&-\delta^{I}_{J}\frac{\partial_{j}}{\nabla^{2}}&-\delta^{I}_{J}\frac{1}{\nabla^{2}}&0&0\\
  \end{array}
\right)\delta^2(x-y).
\label{eq38}
\end{eqnarray}
In this manner, we identify from  (\ref{eq38}) the following   [FJ] brackets
\begin{eqnarray*}
\{e^{I}_{i}(x),\lambda^{J}(y)\}_{FD}&=&\frac{1}{\Lambda}\epsilon_{0ij}\eta^{IJ}\frac{\partial^{j}}{\nabla^{2}}\delta^{2}(x-y), \nonumber\\
\{e^{I}_{i}(x),\rho_{J}(y)\}_{FD}&=&-\delta^{J}_{I}\frac{1}{\nabla^{2}}\delta^{2}(x-y), \nonumber\\
\{A^{I}_{i}(x),\theta^{J}(y)\}_{FD}&=&\epsilon_{0ij}\eta^{IJ}\frac{\partial^{j}}{\nabla^{2}}\delta^{2}(x-y), \nonumber\\
\{A^{I}_{i}(x),\gamma_{J}(y)\}_{FD}&=&-\delta^{J}_{I}\frac{\partial_i}{\nabla^{2}}\delta^{2}(x-y).
\end{eqnarray*}
In particular from  (\ref{eq38}) we also obtain   $\{e^{I}_{i}(x),e^{J}_{j}(y)\}_{FD}=0$ and $\{A^{I}_{i}(x),A^{J}_{j}(y)\}_{FD}=0$, where coincide with the Dirac brackets given in (\ref{134}) and (\ref{137}). However, it is important to comment that by working with the gauge (\ref{25v})   was  not possible  obtain all  Dirac's brackets given from   (\ref{133}) to (\ref{138}) where the first class constraints have  been converted in second class by fixing the gauge. This fact is present   because in Dirac's method  by fixing the gauge we have choosen a particular configuration of the fields, in particular,  the configuration of the canonical momenta, and this fact do not allow   in   [FJ]  framework  to obtain the complete set of  brackets because we used as symplectic variables  the configuration space and the momenta are not invoked, they are labels. In order to obtain  by using [FJ] all Dirac's brackets given from (\ref{133}) to  (\ref{138}), it is necessary to work with the phase space as  symplectic variables, we  will clarify these points in  latter subsection. \\
We finish this section carrying  out the counting of physical degrees of freedom, for this aim we observe that in [FJ] formalism, it is not necessary  to realise   the  classification between the constraints in first class or second class because they are at the same footing. Hence, this fact allow us to carry out the counting of physical degrees of freedom  in a standar way, namely,  there are 12 $(e^{I}_{i},A^{I}_{i})$ canonical variables and there are  12 independent constraints $(\Omega^{(0)}_{I},\beta^{(0)}_{I},\partial^{i}e^{I}_{i},\partial^{i}A^{I}_{i})$, thus, the degrees of freedom=Canonical Variables $-$ constraints$=6-6=0$. In this manner, we conclude that the  abelian exotic action lacks of physical degrees of freedom, i.e., it defines a topological field theory as expected.\\
\subsection{ Faddeev-Jackiw quantization introducing  the phase space  as symplectic variables }
Now, in this section  we will study  the action (\ref{eq2}) by means the [FJ] formalism introducing    the phase space  as symplectic variables. In this manner, from (\ref{eq2}) we identify  the  momenta  $(\pi^{\alpha}_{I},p^{\alpha}_{I})$ canonically conjugate to $(A^{I}_{\alpha},e^{I}_{\alpha})$  given by
\begin{eqnarray}
\pi^{i}_{I}&=&\frac{1}{2}\epsilon^{0ij}A_{jI},\nonumber\\
p^{i}_{I}  &=&\frac{\Lambda}{2}\epsilon^{0ij}e_{jI}.
\label{eq46}
\end{eqnarray}
By using the canonical momenta into the Lagrangian (\ref{eq2}), we obtain the following symplectic Lagrangian
\begin{equation}
\mathcal{L}^{(0)}=\pi^{i}_{I}\dot{A}^{I}_{i}+p^{i}_{I}\dot{e}^{I}_{i}-V^{(0)},
\label{eq48}
\end{equation}
where $V^{(0)} =-2A^{I}_{0}\partial_{i}\pi^{i}_{I}-2e^{I}_{0}\partial_{i}p^{i}_{I}$. In this manner, we can identify  the following set of  symplectic variables  $\xi{^{(0)i}}(x)=\{e^{I}_{i},p^{i}_{I},e^{I}_{0},A^{I}_{i},\pi^{i}_{I},A^{I}_{0}\}$ and the components of the symplectic 1-form   are $a{^{(0)}}{_{i}}(x)=\{p^{i}_{I},0,0,\pi^{i}_{I},0,0\}$. By using these symplectic variables,  the symplectic matrix (\ref{4}) is given by
\begin{eqnarray}
f^{(0)}_{ij}(x,y)=
\left(
 \begin{array}{cccccc}
   0&-\delta{^{i}}_{j}\delta{^{J}}_{I}&0&0&0&0\\
   \delta{^{j}}_{i}\delta{^{I}}_{J}&0&0&0&0&0\\
   0&0&0&0&0&0\\
   0&0&0&0&-\delta{^{i}}_{j}\delta{^{J}}_{I}&0\\
   0&0&0&\delta{^{i}}_{j}\delta{^{I}}_{J}&0&0\\
   0&0&0&0&0&0\\
  \end{array}
\right)\delta^2(x-y).
\label{eq53}
\end{eqnarray}
We realise  that this matrix is singular and thus   the  system has constraints. In order to obtain these constraints,  we calculate the modes of the matrix (\ref{eq53}) given by   $(v_{i}^{(0)})_{1}^{T}=(0,0,v^{e^{I}_{0}},0,0,0)$ and $(v_{i}^{(0)})_{2}^{T}=(0,0,0,0,0,v^{A^{I}_{0}})$, where $v^{e^{I}_{0}}$ and $v^{A^{I}_{0}}$ are   arbitrary functions. In this manner, just like we performed the [FJ] analysis in last section,  by using these modes we obtain the following constraints
\begin{eqnarray}
\Omega^{(0)}_{I}&=&\int d^{2}x(v^{(0)})^{T}_{i}(x)\frac{\delta}{\delta\xi^{(0)i}(x)}\int d^{2}y V^{(0)}(\xi) \nonumber \\
            &=& \int d^{2}x v^{e^{I}_{0}}(x)[-2\partial_{i}p^{i}_{I}] \rightarrow [-2\partial_{i}p^{i}_{I}] =0,
\label{eq54}
\end{eqnarray}
and
\begin{eqnarray}
\Theta^{(0)}_{I}&=&\int d^{2}x(v^{(0)})^{T}_{i}(x)\frac{\delta}{\delta\xi^{(0)i}(x)}\int d^{2}y V^{(0)}(\xi) \nonumber \\
            &=& \int d^{2}x v^{A_{0}}(x)[-2\partial_{i}\pi^{i}_{I}] \rightarrow [-2\partial_{i}\pi^{i}_{I}] =0.
\label{eq55}
\end{eqnarray}
We can observe that these constraints are  the secondary constraints given in (\ref{103a}) and obtained by mean Dirac's method. In order to find out more constraints, we  form the following matrix \cite{8}
\begin{eqnarray}
   f^{}_{kj}\dot{\xi}^{j}=Z_{k}(\xi),
\label{eq58}
\end{eqnarray}
where
\begin{eqnarray}
Z_{k}(\xi)=
\left(
 \begin{array}{cccc}
   \frac{\partial V^{(0)}(\xi)}{\partial \xi^{i}}\\
   0\\
   0\\
  \end{array}
\right),
\label{eq60}
\end{eqnarray}
 and
\begin{eqnarray}
f^{}_{kj}=
\left(
 \begin{array}{cccc}
   f^{(0)}_{ij}\\
   \frac{\partial\Omega^{(0)}}{\partial\xi^{i}}\\
    \frac{\partial\Theta^{(0)}}{\partial\xi^{i}}\\
  \end{array}
\right)=\left( \begin{array}{cccccc}
   0&-\delta{^{i}}_{j}\delta{^{J}}_{I}&0&0&0&0\\
   \delta{^{j}}_{i}\delta{^{I}}_{J}&0&0&0&0&0\\
   0&0&0&0&0&0\\
   0&0&0&0&-\delta{^{i}}_{j}\delta{^{J}}_{I}&0\\
   0&0&0&\delta{^{j}}_{i}\delta{^{J}}_{I}&0&0\\
   0&0&0&0&0&0\\
   0&2\partial_{j}\delta{^{J}}_{I}&0&0&0&0\\
     0&0&0&0&2\partial_{j}\delta{^{J}}_{I}&0\\
  \end{array}
\right)\delta(x-y).
\label{eq59}
\end{eqnarray}
The  matrix (\ref{eq59}) is obviously not a square matrix as expected, but it still has  linearly independent modes, these modes are given by $(v^{(1)})_{1}^{T}=(2\partial_{i}v^{\lambda},0,v^{e^{I}_{0}},0,0,0,v^{\lambda},0)$ and $(v^{(1)})_{2}^{T}=(0,0,0,2\partial_{i}v^{\alpha},0,v^{A^{I}_{0}},0,v^{\alpha})$.  Furthermore, the contraction of the modes $(v^{(1)})_{k}$ with (\ref{eq60}) will lead to  more constraints, this is
\begin{equation}
(v^{(1)})_{k}^{T}Z_{k}\mid_{\Omega^{(0)},\Theta^{(0)}_{I} =0}=0,
\label{eq62}
\end{equation}
with $k=1,2$. It is easy to observe that (\ref{eq62})  corresponds to an identity, therefore, in [FJ] method there are not more constraints for the theory under study.\\
In order to construct a new symplectic Lagrangian containing the information obtained above, we introduce  the Lagrangian multipliers $\lambda^{I}$ and $\rho^{I}$  associated to the constraints, this is
\begin{equation}
\mathcal{L}^{(1)}=\pi^{i}_{I}\dot{A}^{I}_{i}+p^{i}_{I}\dot{e}^{I}_{i}+\Omega^{(0)}_{I}\dot{\lambda}^{I}+\Theta^{(0)}_{I}\dot{\rho}^{I}
\label{eq65}
\end{equation}
where the symplectic potential  $V^{(1)}=V^{(0)}\mid_{\Omega^{(0)}_{I},\Theta^{(0)}_{I}=0}=0$ vanish. Now, from the symplectic Lagrangian (\ref{eq65}) we can identify the following symplectic variables $\xi{^{(1)i}}(x)=\{e^{I}_{i},p^{i}_{I},\lambda^{I},A^{I}_{i},\pi^{i}_{I},\rho^{I}\}$ and the symplectic 1-forms  $a{^{(1)}}{_{i}}(x)=\{p^{i}_{I},0,\partial_{i}p^{i}_{I},\pi^{i}_{I}, 0,\partial_{i}\pi^{i}_{I}\}$. By using these symplectic variables, we obtain the following symplectic matrix
\begin{eqnarray}
f^{(1)}_{ij}(x,y)=
\left(
 \begin{array}{cccccc}
   0&-\delta{^{i}}_{j}\delta{^{J}}_{I}&0&0&0&0\\
   \delta{^{j}}_{i}\delta{^{I}}_{J}&0&-2\delta{^{I}}_{J}\partial_{i}&0&0&0\\
   0&-2\delta{^{J}}_{I}\partial_{j}&0&0&0&0\\
   0&0&0&0&-\delta{^{i}}_{j}\delta{^{J}}_{I}&0\\
   0&0&0&\delta{^{j}}_{i}\delta{^{I}}_{J}&0&-2\delta{^{I}}_{J}\partial_{i}\\
   0&0&0&0&-2\delta{^{J}}_{I}\partial_{j}&0\\
  \end{array}
\right)\delta^2(x-y),
\label{eq67}
\end{eqnarray}
hence,  $f^{(1)}_{ij}$ is a singular matrix, however, we have proved that there are not more constraints and the noninvertibility of (\ref{eq67}) means that the theory has a gauge symmetry.  In order to make invertible  the matrix (\ref{eq67}),  it is necessary  fix  the following  gauge $\partial^{i}e^{I}_{i}=0$ and  $\partial^{i}A^{I}_{i}=0$. In this manner, we introduce new Lagrange multipliers,  namely  $\phi_{I}$ and $\theta_{I}$, associated to the gauge fixing   for constructing the following symplectic Lagrangian
\begin{equation}
\mathcal{L}^{(2)}=\pi^{i}_{I}\dot{A}^{I}_{i}+p^{i}_{I}\dot{e}^{I}_{i}+(2\partial_{i}p^{i}_{I})\dot{\lambda}^{I}+(2\partial_{i}\pi^{i}_{I})\dot{\rho}^{I}+(\partial^{i}e^{I}_{i})\dot{\phi_{I}}+(\partial^{i}A^{I}_{i})\dot{\theta_{I}}.
\label{eq68}
\end{equation}
Hence, from (\ref{eq68}) we identify the following symplectic variables $\xi{^{(2)i}}(x)=\{e^{I}_{i},p^{i}_{I},\lambda^{I},A^{I}_{i},\pi^{i}_{I},\rho^{I},\phi_{I},\theta_{I}\}$ and the symplectic 1-form  $a{^{(2)}}{_{i}}(x)=\{p^{i}_{I},0,2\partial_{i}p^{i}_{I},\pi^{i}_{I},0,2\partial_{i}\pi^{i}_{I},\partial^{i}e^{I}_{i},\partial^{i}A^{I}_{i}\}$. In this manner, by using these symplectic variables, we obtain the following symplectic matrix
\begin{eqnarray}
f^{(2)}_{ij}(x,y)=
\left(
 \begin{array}{cccccccc}
   0&-\delta{^{i}}_{j}\delta{^{J}}_{I}&0&0&0&0&-\delta{^{J}}_{I}\partial_{i}&0\\
   \delta{^{j}}_{i}\delta{^{I}}_{J}&0&-2\delta{^{I}}_{J}\partial_{i}&0&0&0&0&0\\
   0&-2\delta{^{J}}_{I}\partial_{j}&0&0&0&0&0&0\\
   0&0&0&0&-\delta{^{i}}_{j}\delta{^{J}}_{I}&0&0&-\delta{^{J}}_{I}\partial_{i}\\
   0&0&0&\delta{^{j}}_{i}\delta{^{I}}_{J}&0&-2\delta{^{I}}_{J}\partial_{i}&0&0\\
   0&0&0&0&-2\delta{^{J}}_{I}\partial_{j}&0&0&0\\
    -\delta{^{I}}_{J}\partial_{j}&0&0&0&0&0&0&0\\
   0&0&0&-\delta{^{I}}_{J}\partial_{j}&0&0&0&0\\
  \end{array}
\right)\delta^2(x-y).
\label{eq70}
\end{eqnarray}
We can observe that  this  matrix is not singular. Its inverse is given by
\begin{eqnarray*}
[f^{(2)}_{ij}(x,y)]^{-1}=\nonumber\\
\end{eqnarray*}

{\tiny
\begin{eqnarray}
\left(
 \begin{array}{cccccccc}
   0&\delta{^{I}}_{J}(\delta{^{j}}_{i}-\frac{\partial_{i}\partial^{j}}{\nabla^{2}})&0&0&0&0&-\delta{^{I}}_{J}\frac{\partial_{i}}{\nabla^{2}}&0\\
   -\delta{^{J}}_{I}(\delta{^{i}}_{j}-\frac{\partial_{j}\partial^{i}}{\nabla^{2}})&0&-\frac{1}{2}\delta{^{J}}_{I}\frac{\partial^{i}}{\nabla^{2}}&0&0&0&0&0\\
   0&-\frac{1}{2}\delta{^{I}}_{J}\frac{\partial^{j}}{\nabla^{2}}&0&0&0&0&-\delta{^{I}}_{J}\frac{1}{2}\frac{1}{\nabla^{2}}&0\\
    0&0&0&0&\delta{^{I}}_{J}(\delta{^{j}}_{i}-\frac{\partial_{i}\partial^{j}}{\nabla^{2}})&0&0&-\delta{^{I}}_{J}\frac{\partial_{i}}{\nabla^{2}}\\
   0&0&0&-\delta{^{J}}_{I}(\delta{^{i}}_{j}-\frac{\partial_{j}\partial^{i}}{\nabla^{2}})&0&-\frac{1}{2}\delta{^{J}}_{I}\frac{\partial^{i}}{\nabla^{2}}&0&0\\
   0&0&0&0&-\frac{1}{2}\delta{^{I}}_{J}\frac{\partial^{j}}{\nabla^{2}}&0&0&-\frac{1}{2}\delta{^{I}}_{J}\frac{1}{\nabla^{2}}\\
    -\delta{^{J}}_{I}\frac{\partial_{j}}{\nabla^{2}}&0&\frac{1}{2}\delta{^{J}}_{I}\frac{1}{\nabla^{2}}&0&0&0&0&0\\
   0&0&0&-\delta{^{J}}_{I}\frac{\partial_{j}}{\nabla^{2}}&0&\frac{1}{2}\delta{^{J}}_{I}\frac{1}{\nabla^{2}}&0&0\\
  \end{array}
\right) \delta(x-y),
\label{eq71}
\end{eqnarray}}
from which, it is possible   identify the  generalized [FJ] brackets 
\begin{eqnarray}
\{\xi_{i}^{(2)}(x),\xi_{j}^{(2)}(y)\}_{FD}=[f^{(2)}_{ij}(x,y)]^{-1}.
\label{eq72}
\end{eqnarray}
Therefore we find the following brackets
\begin{eqnarray}
\{e^{I}_{i}(x),p^{j}_{J}(y)\}_{FD}=\delta{^{I}}_{J}(\delta{^{j}}_{i}-\frac{\partial_{i}\partial^{j}}{\nabla^{2}})\delta(x-y),
\label{eq74}
\end{eqnarray}
\begin{eqnarray}
\{A^{I}_{i}(x),\pi^{j}_{J}(y)\}_{FD}=\delta{^{I}}_{J}(\delta{^{j}}_{i}-\frac{\partial_{i}\partial^{j}}{\nabla^{2}})\delta(x-y),
\label{eq75}
\end{eqnarray}
\begin{eqnarray}
\{e^{I}_{i}(x),e^{J}_{j}(y)\}_{FD}=0,
\label{eq76}
\end{eqnarray}
\begin{eqnarray}
\{A^{I}_{i}(x),A^{J}_{j}(y)\}_{FD}=0,
\label{eq77}
\end{eqnarray}
\begin{eqnarray}
\{p^{i}_{I}(x),p^{j}_{J}(y)\}_{FD}=0,
\label{eq78}
\end{eqnarray}
\begin{eqnarray}
\{\pi^{i}_{I}(x),\pi^{j}_{J}(y)\}_{FD}=0,
\label{eq79}
\end{eqnarray}
\begin{eqnarray}
\{p^{i}_{I}(x),\lambda^{J}(y)\}_{FD}=-\frac{1}{2}\delta^{J}_{I}\frac{\partial^{i}}{\nabla^{2}}\delta(x-y),
\label{eq80}
\end{eqnarray}
\begin{eqnarray}
\{\pi^{i}_{I}(x),\rho^{J}(y)\}_{FD}=-\frac{1}{2}\delta^{J}_{I}\frac{\partial^{i}}{\nabla^{2}}\delta(x-y),
\label{eq81}
\end{eqnarray}
\begin{eqnarray}
\{\lambda^{I}(x),\phi_{J}(y)\}_{FD}=-\frac{1}{2}\delta^{I}_{J}\frac{1}{\nabla^{2}}\delta(x-y),
\label{eq82}
\end{eqnarray}
\begin{eqnarray}
\{\rho^{I}(x),\theta_{J}(y)\}_{FD}=\frac{1}{2}\delta^{I}_{J}\frac{1}{\nabla^{2}}\delta(x-y),
\label{eq83}
\end{eqnarray}
\begin{eqnarray}
\{e^{I}_{i}(x),\phi_{J}(y)\}_{FD}=-\delta^{I}_{J}\frac{\partial_{i}}{\nabla^{2}}\delta(x-y),
\label{eq84}
\end{eqnarray}
\begin{eqnarray}
\{A^{I}_{i}(x),\theta_{J}(y)\}_{FD}=-\delta^{I}_{J}\frac{\partial_{i}}{\nabla^{2}}\delta(x-y).
\label{eq85}
\end{eqnarray}
In this manner, we can observe that the generalized [FJ] brackets coincide with the Dirac ones obtained in the appendix A (see subsection B)  expressed from   (\ref{133}) to (\ref{138}). Therefore, we finish this section with some comments. We have reproduced by means of  a different way the results obtained by using the  Dirac method applied to the Abelian exotic action. In particular, the [FJ] formalism allowed us  to  obtain the constraints of the theory, the gauge transformations and we have carried out the counting of physical degrees of freedom,   we have also showed  that  if in Dirac's framework  we fix  or not the gauge and we construct the Dirac's brackets,  then the generalized  [FJ] brackets coincide to each other. In the following section we will perform the [FJ] analysis for the non-Abelian theory,  and we will reproduce by means a different and economic way the results reported in \cite{9}.
\section{Faddeev-Jackiw quantization for an  exotic action for gravity  }
Now, we will extend the results obtained in previous  sections by analysing the non-Abelian  action. In particular, we will reproduce the results reported in \cite{9} where a pure Dirac's analysis was performed. In this respect, in \cite{9} was reported the complete structure of the constraints, the Dirac  brackets  were constructed by eliminating only the second class constraints  and  also  all those results were compared with the results obtained by means of the canonical covariant analysis.  Hence, in this section we will obtain the results reported in \cite{9} by means of  [FJ] approach. In order to archive this aim,  we have seen above   that if Dirac's brackets are constructed by eliminating only the second class constraints, then in [FJ]  it is necessary to  work with the configuration space as symplectic variables. In fact, from  (\ref{1}) we can identify the following symplectic Lagrangian
\begin{eqnarray}
\mathcal{L}^{(0)}=\epsilon^{0ij}\eta_{IJ}A^{I}_{j}\partial_{0}A^{J}_{i}+\Lambda\epsilon^{0ij}\eta_{IJ} e^{I}_{j}\partial_{0}e^{i}_{J}-V^{(0)},
\label{eq97}
\end{eqnarray}
where the symplectic potential is given by $V^{(0)}=-2\epsilon^{0ij}\eta_{IJ}\left[F^{J}{_{ij}}+\frac{\Lambda}{2}\epsilon{^{J}}_{KL}e^{K}_{i}e^{L}_{j}\right]A{_{0}}^{I}-2\Lambda\epsilon^{0ij}\eta_{IJ}D_{i}e{_{j}}^{I}e^{J}_{0}$. Thus, from (\ref{eq97}) we identify the following symplectic variables   $\xi{^{(0)i}}(x)=\{e^{I}_{i},e^{I}_{0},A^{I}_{i},A^{I}_{0}\}$ and the symplectic 1-form   $a{^{(0)}}{_{i}}(x)=\{\Lambda\epsilon^{0ij}e_{Ij},0,\epsilon^{0ij}A_{Ij},0\}$. In this manner, by using these symplectic variables, we find that the symplectic matrix has the form
\begin{eqnarray}
f^{(0)}_{ij}(x,y)=
\left(
 \begin{array}{cccc}
   -2\Lambda\epsilon^{0ij}\eta_{IJ}&0&0&0\\
   0&0&0&0\\
   0&0&-2\epsilon^{0ij}\eta_{IJ}&0\\
   0&0&0&0\\
  \end{array}
\right)\delta^2(x-y),
\label{eq99}
\end{eqnarray}
where  we can observe that matrix is singular, this means that there are constraints. We calculate the modes of this matrix; these  modes are given by  $\widetilde{v}^{(0)}_{k}=(0,v^{e^{I}_{0}}(x),0,0)$ and  $\widetilde{w}^{(0)}_{k}=(0,0,0,w^{A^{I}_{0}}(x))$,  where $v^{A^{I}_{0}}$ and $v^{e^{I}_{0}}$ are arbitrary functions. So, just like was  performed in previous sections, we will contract the null vector with the variation of the symplectic potential in order to obtain the constraints, this is
\begin{eqnarray}
\Omega^{(0)}_{I}&=&\int d^{2}x(\widetilde{v}^{(0)})^{T}_{i}(x)\frac{\delta}{\delta\xi^{(0)i}(x)}\int d^{2}y V^{(0)}(\xi) \nonumber \\
            &=& \int d^{2}x v^{e^{I}_{0}}(x)\bigg[-2\Lambda\epsilon^{0ij}\eta_{IJ}D_{i}e{_{j}}^{J}\bigg] \rightarrow \bigg[-2\Lambda\epsilon^{0ij}\eta_{IJ}D_{i}e{_{j}}^{J}\bigg] =0, \nonumber \\
\beta^{(0)}_{I}&=&\int d^{2}x(\widetilde{w}^{(0)})^{T}_{i}(x)\frac{\delta}{\delta\xi^{(0)i}(x)}\int d^{2}y V^{(0)}(\xi) \nonumber \\
            &=& \int d^{2}x w^{A_{0}}(x)\bigg[-2\epsilon^{0ij}\eta_{IJ}\left[F^{J}{_{ij}}+\frac{\Lambda}{2}\epsilon{^{J}}_{KL}e^{K}_{i}e^{L}_{j}\right]\bigg] \rightarrow \bigg[-2\epsilon^{0ij}\eta_{IJ}\left[F^{J}{_{ij}}+\frac{\Lambda}{2}\epsilon{^{J}}_{KL}e^{K}_{i}e^{L}_{j}\right]\bigg] =0,  \nonumber
\label{eq100}
\end{eqnarray}
thus we identify the following constraints
\begin{eqnarray*}
\Omega^{(0)}_{I}=2\Lambda\epsilon^{0ij}\eta_{IJ}D_{i}e{_{j}}^{J}=0,
\end{eqnarray*}
\begin{eqnarray*}
\beta^{(0)}_{I}=2\epsilon^{0ij}\eta_{IJ}\left[F^{J}{_{ij}}+\frac{\Lambda}{2}\epsilon{^{J}}_{KL}e^{K}_{i}e^{L}_{j}\right]=0,
\end{eqnarray*}
these constraints are the secondary constraints found by means Dirac's method and reported in \cite{9}. It is easy to prove that for this theory there are not more [FJ] constraints. In fact, the matrix
\begin{eqnarray}
   f^{(1)}_{kj}\dot{\xi}^{j}=Z_{k}(\xi),
\label{eq58p}
\end{eqnarray}
has the following modes
\begin{eqnarray}
(v^{(1)})_{1}^{T}=(\partial_{i}\textit{v}^{\lambda}+\epsilon{^{I}}_{JK}A^{J}_{i}\textit{v}^{\lambda}+\epsilon{^{I}}_{JK}e^{K}_{i}\textit{v}^{\beta},\textit{v}^{e^{I}_{0}},0,0,\textit{v}^{\lambda},\textit{v}^{\beta}),\nonumber\\
(v^{(1)})_{2}^{T}=(0,0,\partial_{i}\textit{v}^{\beta}+\epsilon{^{I}}_{JK}A^{J}_{i}\textit{v}^{\beta}+\Lambda\epsilon{^{I}}_{JK}e^{K}_{i}\textit{v}^{\lambda},\textit{v}^{A^{I}_{0}},\textit{v}^{\lambda},\textit{v}^{\beta}),
\label{58m}
\end{eqnarray}
and the contraction of these modes with $Z_k$ yield identities, therefore, there are not more constraints. \\
By following with the method,   we will introduce  all this information into the symplectic  Lagrangian  in order to construct a new one,  thus, we introduce the  Lagrangian multipliers $\lambda^I$,  $\theta^I $ associated with  the constraints  $\Omega^{(0)}_{I}$ and $\beta^{(0)}_{I}$ respectively. In this manner, the new symplectic Lagrangian is given by
\begin{eqnarray}
\mathcal{L}^{(1)}=\epsilon^{0ij}\eta_{IJ}A^{I}_{j}\partial_{0}A^{J}_{i}+\Lambda\epsilon^{0ij}\eta_{IJ} e^{I}_{j}\partial_{0}e^{i}_{J}+(\Omega^{(0)}_{I})\dot{\lambda}^{I}+(\beta^{(0)}_{I})\dot{\theta}^{I},
\label{eq101}
\end{eqnarray}
where we can observe that the symplectic potential  vanishes  $V^{(1)}=V^{(0)}\mid_{\Omega^{(0)}_{I}=0,\beta^{(0)}_{I}=0}=0$,  this is an expected result, reflecting  the general covariance of the theory just like it is present in  General Relativity.\\
Now, from (\ref{eq101}) we identify the new set of symplectic variables $ \xi{^{(1)i}}(x)=\{e^{I}_{i},\lambda^{I},A^{I}_{i},\theta^{I},\}$ and the symplectic 1-forms $a{^{(1)}}{_{i}}(x)=\{\Lambda\epsilon^{0ij}e_{Ij},\Omega^{(0)}_{I},\epsilon^{0ij}A_{Ij},\beta^{(0)}_{I}\}$. By using the symplectic variables we can calculate the following symplectic matrix
\begin{eqnarray}
f^{(1)}_{ij}(x,y)=\nonumber
\end{eqnarray}

\begin{eqnarray}
{\tiny
\left(
 \begin{array}{cccc}
   -2\Lambda\epsilon^{0ij}\eta_{IJ}&-2\Lambda\epsilon^{0ij}(\eta_{IJ}\partial_{j}+\epsilon_{IJK}A^{K}_{j})&0&-2\Lambda\epsilon^{0ij}\epsilon_{IJK}e^{K}_{j}\\
   2\Lambda\epsilon^{0ji}(\eta_{IJ}\partial_{i}-\epsilon_{IJK}A^{K}_{i})&0&-2\Lambda\epsilon^{0ji}\epsilon_{IJK}e^{K}_{i}&0\\
   0&-2\Lambda\epsilon^{0ij}\epsilon_{IJK}e^{K}_{j}&-2\epsilon^{0ij}\eta_{IJ}&-2\epsilon^{0ij}(\eta_{IJ}\partial_{j}+\epsilon_{IJK}A^{K}_{j})\\
  -2\Lambda\epsilon^{0ji}\epsilon_{IJK}e^{K}_{i}&0&2\epsilon^{0ji}(\eta_{IJ}\partial_{i}-\epsilon_{IJK}A^{K}_{i})&0\\
  \end{array}
\right)\delta^2(x-y)}, \nonumber \\
\label{eq103}
\end{eqnarray}
we can observe that  $f^{(1)}_{ij}$ is singular, however, we have commented that there are not more constraints; the noninvertibility of (\ref{eq103})  indicate  that the theory has a gauge symmetry. Hence, we choose the following gauge fixing as constraints
\begin{eqnarray}
A^{I}_{0}(x)&=&0,  \nonumber\\
e^{I}_{0}(x)&=&0, \nonumber
\end{eqnarray}
then we  introduce the   Lagrangians multipliers $\phi_{I}$ and $\alpha_{I}$ associated with the above gauge fixing for constructing  a new  symplectic Lagrangian
\begin{eqnarray}
\mathcal{L}^{(2)}=\epsilon^{0ij}\eta_{IJ}A^{I}_{j}\partial_{0}A^{J}_{i}+\Lambda\epsilon^{0ij}\eta_{IJ} e^{I}_{j}\partial_{0}e^{i}_{J}+(\Omega^{(0)}_{I}+\phi_{I})\dot{\lambda}^{I}+(\beta^{(0)}_{I}+\alpha_{I})\dot{\theta}^{I},
\label{eq104}
\end{eqnarray}
thus, we identify the following set of symplectic  variables
$\xi{^{(2)i}}(x)=\{e^{I}_{i},\lambda^{I},A^{I}_{i},\theta^{I},\phi_{I},\alpha_{I}\}$  and the symplectic 1-forms $a{^{(2)}}{_{i}}(x)=\{\Lambda\epsilon^{0ij}e_{Ij},\Omega^{(0)}_{I}+\phi_{I},\epsilon^{0ij}A_{Ij},\beta^{(0)}_{I}+\alpha_{I},0,0\}$. Furthermore, by using these symplectic variables we find that the symplectic matrix is given by
\begin{eqnarray}
f^{(2)}_{ij}(x,y)=\nonumber
\end{eqnarray}
\begin{eqnarray}
{\tiny
\left(
 \begin{array}{cccccc}
   -2\Lambda\epsilon^{0ij}\eta_{IJ}&-2\Lambda\epsilon^{0ij}(\eta_{IJ}\partial_{j}+\epsilon_{IJK}A^{K}_{j})&0&-2\Lambda\epsilon^{0ij}\epsilon_{IJK}e^{K}_{j}&0&0\\
   2\Lambda\epsilon^{0ji}(\eta_{IJ}\partial_{i}-\epsilon_{IJK}A^{K}_{i})&0&-2\Lambda\epsilon^{0ji}\epsilon_{IJK}e^{K}_{i}&0&-\delta^{J}_{I}&0\\
   0&-2\Lambda\epsilon^{0ij}\epsilon_{IJK}e^{K}_{j}&-2\epsilon^{0ij}\eta_{IJ}&-2\epsilon^{0ij}(\eta_{IJ}\partial_{j}+\epsilon_{IJK}A^{K}_{j})&0&0\\
  -2\Lambda\epsilon^{0ji}\epsilon_{IJK}e^{K}_{i}&0&2\epsilon^{0ji}(\eta_{IJ}\partial_{i}-\epsilon_{IJK}A^{K}_{i})&0&0&-\delta^{J}_{I}\\
   0&\delta^{I}_{J}&0&0&0&0\\
  0&0&0&\delta^{I}_{J}&0&0
  \end{array}
\right)\delta^2(x-y)}, \nonumber \\
\label{eq106}
\end{eqnarray}
we observe that $f^{(2)}_{ij}$ is not singular, hence, it is an invertible matrix. After a long calculation, the inverse is given by
\begin{eqnarray}
[f^{(2)}_{ij}(x,y)]^{-1}=\nonumber
\end{eqnarray}
\begin{eqnarray}
{\tiny
\left(
 \begin{array}{cccccc}
   \frac{1}{2\Lambda}\epsilon_{0ij}\eta^{IJ}&0&0&0&-(\delta^{I}_{J}\partial_{i}+\epsilon{^{I}}_{JK}A^{K}_{i})&-\epsilon^{0ij}\epsilon{^{I}}_{JK}e^{K}_{i}\\
   0&0&0&0&\delta^{J}_{I}&0\\
0&0&\epsilon_{0ij}\frac{1}{2}\eta^{IJ}&0&-\Lambda\epsilon{^{I}}_{JK}e^{K}_{i}&-(\delta^{I}_{J}\partial_{i}+\epsilon{^{I}}_{JK}A^{K}_{i})\\
   0&0&0&0&0&\delta^{J}_{I}\\
   (\delta^{J}_{I}\partial_{j}-\epsilon_{I}{^{J}}_{K}A^{K}_{j})&-\delta^{I}_{J}&0&-\Lambda\epsilon_{I}{^{J}}_{K}e^{K}_{j}&0&0\\
 -\epsilon_{I}{^{J}}_{K}e^{K}_{j}&0&(\delta^{J}_{I}\partial_{j}-\epsilon_{I}{^{J}}_{K}A^{K}_{j})&-\delta^{I}_{J}&0&0\\
  \end{array}
\right)\delta^2(x-y)}.
\label{eq107}
\end{eqnarray}
Therefore, from (\ref{eq107}) it is possible to identify the following [FJ] generalized brackets given by
\begin{eqnarray}
\{e^{I}_{i}(x),e^{J}_{j}(y)\}_{FD}&=&\frac{1}{2\Lambda}\epsilon_{0ij}\eta^{IJ}\delta^{2}(x-y),\nonumber\\
\{A^{I}_{i}(x),A^{J}_{j}(y)\}_{FD}&=&\frac{1}{2}\epsilon_{0ij}\eta^{IJ}\delta^{2}(x-y),\nonumber\\
\{e^{I}_{i}(x),\phi_{J}(y)\}_{FD}&=&(\delta^{I}_{J}\partial_{i}-\epsilon{^{I}}_{JK}A^{K}_{i})\delta^{2}(x-y),\nonumber\\
\{A^{I}_{i}(x),\alpha_{J}(y)\}_{FD}&=&(\delta^{I}_{J}\partial_{i}-\epsilon{^{I}}_{JK}A^{K}_{i})\delta^{2}(x-y),\nonumber\\
\{e^{I}_{i}(x),\alpha_{J}(y)\}_{FD}&=&-\epsilon{^{I}}_{JK}e^{K}_{i}\delta^{2}(x-y),\nonumber\\
\{A^{I}_{i}(x),\phi_{J}(y)\}_{FD}&=&-\Lambda\epsilon{^{I}}_{JK}e^{K}_{i}\delta^{2}(x-y),\nonumber\\
\{\lambda^{I}(x),\phi_{J}(y)\}_{FD}&=&\delta^{J}_{I}\delta^{2}(x-y),\nonumber\\
\{\theta^{I}(x),\alpha_{J}(y)\}_{FD}&=&\delta^{J}_{I}\delta^{2}(x-y).
\label{eq108}
\end{eqnarray}
It is important to comment,  that the generalized [FJ] brackets coincide with those obtained by means of  the  Dirac method reported in \cite{9}. In fact,  if we make a  redefinition of the fields introducing the momenta 
\begin{eqnarray}
p^{i}_{I}&=&\Lambda\epsilon^{0ij}\eta_{IJ}e^{J}_{j},\nonumber\\
\pi^{i}_{I}&=&\epsilon^{0ij}\eta_{IJ}A^{J}_{j},
\end{eqnarray}
the generalized [FJ] brackets (\ref{eq108}) take the form
\begin{eqnarray}
\{e^{I}_{i}(x),e^{J}_{j}(y)\}_{FD}&=&\frac{1}{2\Lambda}\epsilon_{0ij}\eta^{IJ}\delta^{2}(x-y),\nonumber\\
\{A^{I}_{i}(x),A^{J}_{j}(y)\}_{FD}&=&\frac{1}{2}\epsilon_{0ij}\eta^{IJ}\delta^{2}(x-y),\nonumber\\
\{e^{I}_{i}(x),p^{j}_{J}(y)\}_{FD}&=&\frac{1}{2}\delta^{j}_{i}\delta^{I}_{J}\delta^{2}(x-y),\nonumber\\
\{A^{I}_{i}(x),\pi^{j}_{J}(y)\}_{FD}&=&\frac{1}{2}\delta^{j}_{i}\delta^{I}_{J}\delta^{2}(x-y),\nonumber\\
\{p^{i}_{J}(x),p^{j}_{J}(y)\}_{FD}&=&\frac{\Lambda}{2}\epsilon^{0ij}\eta_{IJ}\delta^{2}(x-y),\nonumber\\
\{\pi^{i}_{I}(x),\pi^{j}_{J}(y)\}_{FD}&=&\frac{1}{2}\epsilon^{0ij}\eta_{IJ}\delta^{2}(x-y),
\end{eqnarray}
which  coincide with  the full Dirac's brackets reported in \cite{9}.\\
Furthermore, with all  constraints at hand, we can carryout the counting of physical degrees of freedom in the following form; there are $12$  dynamical variables $(e^I_i, A^I_i)$  and  $12$ constraints $(\Omega^{(0)}_{I} , \beta^{(0)}_{I}, A^{I}_{0}, e^{I}_{0} )$ , therefore, the theory lacks of physical degrees of freedom.\\
We finish this section by calculating the gauge transformations of the theory. For this aim we calculate the modes of the matrix (\ref{eq103}), those modes are given by 
\begin{eqnarray*}
(w^{(1)})_{1}^{T}=(\partial_{i}\varepsilon^{I}+\epsilon{^{I}}_{JK}A^{J}_{i}\varepsilon^{K}+\epsilon{^{I}}_{JK}e^{K}_{i}\zeta^{J},\varepsilon^{I},0,\zeta^{I}),\nonumber\\
(w^{(1)})_{2}^{T}=(0,\varepsilon^{I},\partial_{i}\zeta^{I}+\epsilon{^{I}}_{JK}A^{J}_{i}\zeta^{K}+\Lambda\epsilon{^{I}}_{JK}e^{K}_{i}\varepsilon^{J},\zeta^{I}).\nonumber
\end{eqnarray*}
In agreement with the [FJ] symplectic formalism, the zero-modes $(w^{(1)})_{1}^{T}$ and $(w^{(1)})_{2}^{T}$ are the generators of infinitesimal gauge transformations of the action (\ref{eq97})  and are given by

\begin{eqnarray*}
\delta e^{I}_{i}(x)&=&D_{i}\varepsilon^{I}+\epsilon{^{I}}_{JK}e^{K}_{i}\zeta^{J} ,\nonumber \\
\delta e^{I}_{0}(x)&=&\partial_{0}\varepsilon^{I}, \nonumber \\
\delta A^{I}_{i}(x)&=&D_{i}\zeta^{I}+\Lambda\epsilon{^{I}}_{JK}e^{K}_{i}\varepsilon^{J}, \nonumber \\
\delta A^{I}_{0}(x)&=&\partial_{0}\zeta^{I}.\nonumber
\end{eqnarray*}
In this manner, by using the   [FJ] symplectic framework we have reproduced the fundamental gauge transformations corresponding to a  $\Lambda$-deformed $ISO(2, 1)$ Poincar\'e transformations reported in \cite{9}.
\section{ Conclusions}
In this paper a detailed  Hamiltonian and [FJ]  analysis for an Abelian exotic action and for a non-Abelian exotic action for gravity in three dimensions have  been performed. With respect to the Abelian theory,  by using the [FJ] we have found the constraints, the gauge transformation,  we had  carried  out the counting of physical degrees of freedom and we have obtained the generalized [FJ] brackets. We could observe that if we work with the configuration space as  symplectic variables, then  we reproduce  the results found by means the Dirac approach where Dirac's brackets are   constructed by eliminating  only the second class constraints. On the other hand, if  in Dirac's framework we  convert the first class constraints,  in second class constraints by fixing the gauge and we calculate the new Dirac's brackets, then in order to reproduce those results,   in [FJ] method  it is necessary to work with the phase space as symplectic variables.  We showed that  by fixing or not the gauge, the Dirac brackets and generalised [FJ] brackets coincide to each other. It is important to remark, that  if in Dirac's approach we fix the gauge and then we  construct the Dirac brackets,  in the [FJ] scheme we could  not reproduce these results  by working with the configuration space. In fact,  the gauge fixing in   Dirac's method   implies  to take    a particular configuration of the fields and the momenta. In [FJ]  by working with the configuration space,  the momenta are labels, however, if we choose the phase space as symplectic variables and we fix the gauge in order to invert the symplectic matrix, now we are choosing a particular configuration of the fields and the momenta as well, then it is possible  reproduce the results obtained in the Dirac approach by fixing the gauge. \\
Furthermore, in the case of a non-Abelian theory, we obtained by means the  [FJ] method the complete set of constraints, the gauge transformations and we carried  out the counting of physical degrees of freedom. In particular, we have reproduced by means a different and economical way the results reported in \cite{9} where  was performed a pure  Dirac's method. \\
We finish this paper with some comments. We have seen that in [FJ] framework it is not necessary to classify  the constraints in second class  or first class as   in Dirac's method is done, and this fact allows that the [FJ] method is more convenient to perform. In this sense, we can perform the analysis to other models describing three dimensional gravity. In fact, there is an alternative  model reproducing Einstein's equations with a cosmological constant and  given by  \cite{13x}
\begin{eqnarray}
S[A,e]=S'[A,e]+\frac{1}{\gamma}\widetilde{S}[A,e], 
\label{ac69}
\end{eqnarray}
where $S'[A,e]$ is the Palatini action,  $\widetilde{S}[A,e]$ is the exotic action analysed in this work and $\gamma$ is a kind of Barbero-Immirzi parameter. In fact, the Hamiltonian analysis of the action (\ref{ac69}) has been reported in \cite{13x},  in particular, in that work  the Dirac brackets have been constructed only eliminating the second class constraints. In this respect,  by using the results obtained in this work, we can develop  the [FJ] analysis of (\ref{ac69}), in particular we   will report  an  easy   way for calculating  the algebra between  the constraints  by means  [FJ] framework  \cite{17}. \\
Finally, we would to comment that   we have at hand  all the necessary tools for performing  the [FJ] analysis of  theories  with a difficult Hamiltonian structure where there are present  tertiary  constraints just like it is present in  topologically massive gravity  \cite{17a, 18, 19, 20, 21, 22, 23, 24}. In fact,  it is well-known that the canonical analysis of  topologically massive gravity is not easy to perform. In the analysis there are present  primary, secondary  and tertiary  constraints. Furthermore, the  classification of those constraints in second class and first class  is  not an easy work, there are several complications in the computations in order to identify the constraints. In fact,  in topologically massive gravity  there are  physical degrees of freedom, however, because of the hamiltonian analysis is difficult to carry out, there are inconsistencies in the counting of physical degrees of freedom \cite{19}. In this respect, our work could be an important tool for studying those theories in the context of [FJ]. These  ideas are in progress and  will be the subject of forthcoming works.    \\ 
\newline
\newline
\noindent \textbf{Acknowledgements}\\[1ex]
This work was supported by CONACyT under Grant No. CB-2014-01/ 240781. We would like to thank R. Cartas-Fuentevilla for discussion on the subject and reading of the manuscript.
\section{ appendix }
\subsection{Dirac's method for an abelian exotic action}
The action that we will study in this appendix  is given by
\begin{equation}
S[A, e]= \int \frac{1}{2}\epsilon^{\mu\nu\lambda}(A^{I}_{\mu}\partial_{\nu}A_{\lambda I}+\Lambda e^{I}_{\mu}\partial_{\nu}e_{\lambda I}) dx^3,
\label{122}
\end{equation}
here, $A_{\mu}$ is the gauge potential with $\mu=0,1,2$ denoting the space-time components. We adopt the following conventions $\epsilon^{012}= \epsilon_{012}=1$. The equations of motion obtained from  (\ref{122}) are
\begin{equation}
\frac{\delta S[A,e]}{ \delta e_{\mu}^I}: \qquad \epsilon^{\mu\nu\rho}\Lambda\partial_{\nu}e{_{\rho}}^{I}=0,
\label{eq71x}
\end{equation}
\begin{equation}
\frac{\delta S[A,e]}{ \delta A{_{\mu}I}}: \qquad \epsilon^{\mu\nu\rho} \partial_{\nu}A{_{\rho}}^{I}=0.
\end{equation}
We can see that (\ref{eq71x}) implies  $e_{\alpha I}=\partial_\alpha f_I$, thus  $g_{\mu \nu}= \eta_{IJ} \partial_\mu f^I \partial_\nu f^J$, which corresponds to  (locally) Minkowski spacetime. We shall resume the complete Hamiltonian analysis of the action (\ref{122}); for this aim, we perform  the 2 + 1 decomposition   and introducing the canonical momenta $(\pi^{\alpha}_{I},p^{\alpha}_{I})$ canonically conjugate to $(A^{I}_{\alpha},e^{I}_{\alpha})$ given by
\begin{equation}
\pi^{\lambda}_{I}:=\frac{\partial\mathcal{L}}{\partial\dot{A}^{I}_{\lambda}}=\frac{1}{2}\epsilon^{0\lambda\gamma}A_{\gamma I},
\end{equation}
\begin{equation}
p^{\lambda}_{I}:=\frac{\partial\mathcal{L}}{\partial\dot{e}^{I}_{\lambda}}=\frac{\Lambda}{2}\epsilon^{0\lambda\gamma}e_{\gamma I},
\end{equation}
with the following fundamental  Poisson brackets among the canonical variables
\begin{eqnarray}
\{A^{I}_{\mu}(x),\pi^{\nu}_{J}(y)\}=\delta_{\mu}^{\nu}\delta^{I}_{J}\delta^{2}(x-y),\\
\{e^{I}_{\mu}(x),p^{\nu}_{J}(y)\}=\delta_{\mu}^{\nu}\delta^{I}_{J}\delta^{2}(x-y),
\end{eqnarray}
we obtain the following primary constraints
\begin{eqnarray}
\Phi^{0}_{I}&:=& \pi^{0}_{I} \approx 0 ,\nonumber \\
\Phi^{i}_{I}&:=& \pi^{i}_{I}- \frac{1}{2}\epsilon^{0ij}A_{jI}\approx 0,\nonumber \\
\phi^{0}_{I}&:=& p^{0}_{I} \approx 0 ,\nonumber \\
\phi^{i}_{I}&:=& p
^{i}_{I}- \frac{\Lambda}{2}\epsilon^{0ij}e_{jI}\approx 0.
\end{eqnarray}
From consistency of the primary constraints, we obtain the following secondary constraints
\begin{eqnarray}
 \psi_{I}&:=& 2\partial_{i}p^{i}_{I} \approx 0, \nonumber \\
 \theta_{I}&:= &2\partial_{i}\pi^{i}_{I} \approx 0.
 \label{103a}
\end{eqnarray}
For this theory there are no, third constraints. Now, from the primary and secondary constraints,  we need to identify  which ones correspond to first and second class. For this aim,  we need to calculate the rank and the null-vectors of the following $8\times8$ matrix whose entries will be the Poisson brackets between primary and secondary constraints given by
\begin{eqnarray}
\{\phi^{i}_{I}(x),\phi^{j}_{J}(y)\} &=& -\Lambda\epsilon^{0ij}\eta_{IJ}\delta^{2}(x-y), \nonumber \\
\{\phi^{i}_{I}(x),\Phi^{j}_{J}(y)\} &=& 0,\nonumber \\
\{\phi^{i}_{I}(x),\psi_{J}(y)\} &=&\Lambda\epsilon^{0ij}\eta_{IJ}\partial^{}_{j} \delta^2(x-y).\nonumber \\
\{\Phi^{i}_{I}(x),\Phi^{j}_{J}(y)\} &=&-\epsilon^{0ij}\eta_{IJ}\delta^2(x-y).\nonumber \\
\{\Phi^{i}_{I}(x),\theta_{J}(y)\} &=&\epsilon^{0ij}\eta_{IJ}\partial^{}_{j} \delta^2(x-y).
\end{eqnarray}
This matrix has rank=4 and 4 null vectors, this mean that the  theory presents a set of 4 first class constraints and 4 second class constraints. In this manner, by using the null vectors,  we identify the following 4 first class constraints
\begin{eqnarray}
\gamma^{1}_I&=& p^{0}_{I} \approx 0,  \nonumber \\
\gamma^{2}_I&=& 2\partial_ip^{i}_{I}-\partial_{i}\phi^{i}_{I} \approx 0, \nonumber \\
\gamma^{3}_I&=& \pi^{0}_{I} \approx 0, \nonumber \\
\gamma^{4}_I&=& 2\partial_i\pi^{i}_{I}-\partial_{i}\Phi^{i}_{I} \approx 0,
\end{eqnarray}
and the rank allows  us  identify  the following 4 second class constraints
\begin{eqnarray}
\chi^{1}_I&=&p^{i}_{I}-\frac{\Lambda}{2}\epsilon^{0ij}e_{jI} \approx 0,\nonumber \\
\chi^{2}_I&=&\pi^{i}_{I}-\frac{1}{2}\epsilon^{0ij}A_{jI} \approx.
\label{134n}
\end{eqnarray}
A direct calculation leads to  the  following non zero brackets  between the first and second class constraints are
\begin{eqnarray}
\{\chi^{1}(x),\chi^{1}(y) \} &=&-\Lambda\epsilon^{0ij}\eta_{IJ}\delta^{2}(x-y), \nonumber \\
\{\chi^{2} (x),\chi^{2}(y) \} &=&-\epsilon^{0ij}\eta_{IJ}\delta^{2}(x-y), 
\label{134m}
\end{eqnarray}
these Poisson brackets   have the following matrix form
\begin{eqnarray}
C^{ij}=
\left(
 \begin{array}{cccc}
   -\Lambda&0\\
   0&-1\\
  \end{array}
\right)\epsilon^{0ij}\eta_{IJ}\delta^2(x-y),
\end{eqnarray}
and its inverse will be
\begin{eqnarray}
[C^{ij}]{^{-1}}=
\left(
 \begin{array}{cccc}
   -\frac{1}{\Lambda}&1\\
   0&-1\\
  \end{array}
\right)\epsilon_{0ij}\eta^{IJ}\delta^2(x-y).
\label{136m}
\end{eqnarray}
With all these results, we can eliminate the second class constraints by introducing the Dirac brackets. Hence,  the Dirac  brackets among two functionals $A$, $B$  expressed by
\begin{eqnarray}
\{A(x),B(y)\}_{D}=\{A(x),B(y)\}_{P}-\int dudv\{A(x),\zeta^{i}(u)\}[C^{ij}]{^{-1}}(u,v)\{\zeta^{j}(v),B(y)\},
\label{137}
\end{eqnarray}
where $\{A(x),B(y)\}_{P}$ is the usual Poisson brackets between the functionals $A$, $B$,  $\zeta^{i}(u)=(\chi^{1},\chi^{2})$ are the second class constraints and  $C^{ij}{^{-1}}$ is given in (\ref{136m}). In this manner, by using this fact  we  obtain the following Dirac's brackets of the theory
\begin{eqnarray}
\{e^{I}_{i}(x),e^{J}_{j}(y)\}_{D}&=&\frac{1}{\Lambda}\epsilon_{0ij}\eta^{IJ}\delta^{2}(x-y),
\label{125}
\end{eqnarray}
\begin{eqnarray}
\{e^{I}_{i}(x),p^{j}_{J}(y)\}_{D}&=&\frac{1}{2}\delta^{j}_{i}\delta^{I}_{J}\delta^{2}(x-y),
\label{126}
\end{eqnarray}
\begin{eqnarray}
\{p^{i}_{I}(x),p^{j}_{J}(y)\}_{D}&=&\frac{\Lambda}{4}\epsilon^{0ij}\eta_{IJ}\delta^{2}(x-y),
\label{127}
\end{eqnarray}
\begin{eqnarray}
\{A^{I}_{i}(x),A^{J}_{j}(y)\}_{D}&=&\epsilon_{0ij}\eta^{IJ}\delta^{2}(x-y),
\label{128}
\end{eqnarray}
\begin{eqnarray}
\{A^{I}_{i}(x),\pi^{j}_{J}(y)\}_{D}&=&\frac{1}{2}\delta^{j}_{i}\delta^{I}_{J}\delta^{2}(x-y),
\label{129}
\end{eqnarray}
\begin{eqnarray}
\{\pi^{i}_{I}(x),\pi^{j}_{J}(y)\}_{D}&=&\frac{1}{4}\epsilon^{0ij}\eta_{IJ}\delta^{2}(x-y),
\label{130}
\end{eqnarray}
\begin{eqnarray}
\{e^{I}_{0}(x),p^{0}_{J}(y)\}_{D}&=&\delta^{I}_{J}\delta^{2}(x-y),
\label{131}
\end{eqnarray}
\begin{eqnarray}
\{A^{I}_{0}(x),\pi^{0}_{J}(y)\}_{D}&=&\delta^{I}_{J}\delta^{2}(x-y).
\label{132}
\end{eqnarray}
Therefore, in order to quantize the theory, we consider  to the second class constraints (\ref{134n}) as strong identities and Dirac's brackets are promoted to commutator.  It is worth to comment, that the Dirac brackets given above are a particular case of the nonabelian case reported in  \cite{9}.\\
\subsection{By fixing the gauge}
In spite of we have eliminated the second class constraints, it is necessary to  remove all the gauge freedom of the theory. In order to archive this aim,  we need to  impose gauge conditions,  using for example  the  temporal and Coulomb gauge 
\begin{eqnarray}
\Omega_{1}&=&e^{I}_{0}\approx0 ,\nonumber \\
\Omega_{2}&=&\partial^{i}e^{I}_{i}\approx0, \nonumber \\
\Omega_{1}&=&A^{I}_{0}\approx0, \nonumber \\
\Omega_{2}&=&\partial^{i}A^{I}_{i}\approx0,
\end{eqnarray}
thus we obtain the complete set of second class constraints
\begin{eqnarray}
\chi^{1}_I&=& p^{0}_{I} \approx 0,  \nonumber \\
\chi^{2}_I&=& 2\partial_ip^{i}_{I}-\partial_{i}\phi^{i}_{I} \approx 0, \nonumber \\
\chi^{3}_I&=& \pi^{0}_{I} \approx 0, \nonumber \\
\chi^{4}_I&=& 2 \partial_i\pi^{i}_{I}-\partial_{i}\Phi^{i}_{I} \approx 0, \nonumber \\
\chi^{5}_I&=&p^{i}_{I}-\frac{\Lambda}{2}\epsilon^{0ij}e_{jI} \approx 0,\nonumber \\
\chi^{6}_I&=&\pi^{i}_{I}-\frac{1}{2}\epsilon^{0ij}A_{jI}, \nonumber \\
\chi_{7}^I&=&e^{I}_{0}\approx0, \nonumber \\
\chi_{8}^I&=&\partial^{i}e^{I}_{i},\approx0, \nonumber \\
\chi_{9}^I&=&A^{I}_{0}\approx 0, \nonumber \\
\chi_{10}^I&=&\partial^{i}A^{I}_{i}\approx 0.
\end{eqnarray}
Now, in order to construct the new Dirac's brackets  we need to  calculate the matrix $C_{ij}$ whose entries are given by the Poisson brackets between the second class constraints. That matrix has the following form
\begin{eqnarray}
C_{ij}=
\left(
 \begin{array}{cccccccccccc}
 0&-\Lambda\eta_{IJ}&0&0&0&-\delta^{J}_{I}\partial_{1}&0&0&0&0&0&0\\
 \Lambda\eta_{IJ}&0&0&0&0&-\delta^{J}_{I}\partial_{2}&0&0&0&0&0&0\\
 0&0&0&-\delta^{J}_{I}&0&0&0&0&0&0&0&0\\
 0&0&\delta^{I}_{J}&0&0&0&0&0&0&0&0&0\\
  0&0&0&0&0&-\delta^{J}_{I}\nabla^{2}&0&0&0&0&0&0\\
 -\delta^{I}_{J}\partial_{1}&-\delta^{I}_{J}\partial_{2}&0&0&\delta^{I}_{J}\nabla^{2}&0&0&0&0&0&0&0\\
  0&0&0&0&0&0&0&-\delta_{IJ}&0&0&0&-\delta^{J}_{I}\partial_{1}\\
 0&0&0&0&0&0&\delta_{IJ}&0&0&0&0&-\delta^{J}_{I}\partial_{2}\\
  0&0&0&0&0&0&0&0&0&-\delta^{J}_{I}&0&0\\
  0&0&0&0&0&0&0&0&\delta^{I}_{J}&0&0&0\\
   0&0&0&0&0&0&0&0&0&0&0&-\delta^{J}_{I}\nabla^{2}\\
  0&0&0&0&0&0&-\delta^{I}_{J}\partial_{1}&-\delta^{I}_{J}\partial_{2}&0&0&\delta^{I}_{J}\nabla^{2}&0\\
  \end{array}
\right)\delta^2(x-y),\nonumber
\end{eqnarray}
and its inverse is given by
{\tiny
\begin{eqnarray}
[C^{ij}]^{-1}=
\left(
 \begin{array}{cccccccccccc}
 0&\Lambda^{-1}\eta^{IJ}&0&0&-\Lambda^{-1}\eta^{IJ}\frac{\partial_{2}}{\nabla^{2}}&0&0&0&0&0&0&0\\
 -\Lambda^{-1}\eta^{IJ}&0&0&0&\Lambda^{-1}\eta^{IJ}\frac{\partial_{1}}{\nabla^{2}}&0&0&0&0&0&0&0\\
 0&0&0&\delta^{J}_{I}&0&0&0&0&0&0&0&0\\
 0&0&-\delta^{I}_{J}&0&0&0&0&0&0&0&0&0\\
  -\Lambda^{-1}\eta^{IJ}\frac{\partial_{2}}{\nabla^{2}}&\Lambda^{-1}\eta^{IJ}\frac{\partial_{1}}{\nabla^{2}}&0&0&0&\delta^{J}_{I}\frac{1}{\nabla^{2}}&0&0&0&0&0&0\\
 0&0&0&0&-\delta^{I}_{J}\frac{1}{\nabla^{2}}&0&0&0&0&0&0&0\\
 0&0&0&0&0&0&0&\eta^{IJ}&0&0&-\eta^{IJ}\frac{\partial_{2}}{\nabla^{2}}&0\\
 0&0&0&0&0&0&-\eta^{IJ}&0&0&0&\eta^{IJ}\frac{\partial_{1}}{\nabla^{2}}&0\\
 0&0&0&0&0&0&0&0&0&\delta^{J}_{I}&0&0\\
 0&0&0&0&0&0&0&0&-\delta^{I}_{J}&0&0&0\\
 0&0&0&0&0&0&-\eta^{IJ}\frac{\partial_{2}}{\nabla^{2}}&\eta^{IJ}\frac{\partial_{1}}{\nabla^{2}}&0&0&0&\delta^{J}_{I}\frac{1}{\nabla^{2}}\\
 0&0&0&0&0&0&0&0&0&0&-\delta^{I}_{J}\frac{1}{\nabla^{2}}&0\
  \end{array}
\right)\delta^2(x-y), \nonumber
\end{eqnarray}}
In this manner, by using the definition of Dirac's brackets among two functionals, we obtain the following  Dirac's brackets among the fields
\begin{eqnarray}
\{e^{I}_{i}(x),p^{j}_{J}(y)\}_{D}=\delta{^{I}}_{J}(\delta{^{j}}_{i}-\frac{\partial_{i}\partial^{j}}{\nabla^{2}})\delta(x-y),
\label{133}
\end{eqnarray}
\begin{eqnarray}
\{e^{I}_{i}(x),e^{J}_{j}(y)\}_{D}=0,
\label{134}
\end{eqnarray}
\begin{eqnarray}
\{p^{i}_{I}(x),p^{j}_{J}(y)\}_{D}=0,
\label{135}
\end{eqnarray}
\begin{eqnarray}
\{A^{I}_{i}(x),\pi^{j}_{J}(y)\}_{D}=\delta{^{I}}_{J}(\delta{^{j}}_{i}-\frac{\partial_{i}\partial^{j}}{\nabla^{2}})\delta(x-y),
\label{136}
\end{eqnarray}
\begin{eqnarray}
\{A^{I}_{i}(x),A^{J}_{j}(y)\}_{D}=0,
\label{137}
\end{eqnarray}
\begin{eqnarray}
\{\pi^{i}_{I}(x),\pi^{j}_{J}(y)\}_{D}=0.
\label{138}
\end{eqnarray}
We can observe that these brackets coincide with those calculated by means the  [FJ] framework.


\end{document}